\documentclass[aps,prl,preprint,groupedaddress]{revtex4-2}

\usepackage{amsfonts}
\usepackage{graphicx}%
\usepackage{grffile}
\usepackage{threeparttable}
\usepackage{booktabs}%
\usepackage{mathtools}
\usepackage{makecell}
\usepackage{braket}
\usepackage{enumitem}
\begin{document}

\title{Quantum Entanglement as Super-Confounding: From Bell's Theorem to Robust Machine Learning}


\author{Pilsung Kang}
\email{pilsungk@dankook.ac.kr}
\thanks{This work was supported by the National Research Foundation of Korea (NRF) grant funded by the Korea government (Ministry of Science and ICT (MSIT)), grant number 2020R1F1A1067619.}
\thanks{This research was supported by `Quantum Information Science R\&D Ecosystem Creation' through NRF funded by the Korea government (MSIT), grant number 2020M3H3A1110365.}
\affiliation{Department of Software Science, Dankook University\\ Yongin 16890, South Korea}

\date{\today}

\begin{abstract}
Bell's theorem reveals a profound conflict between quantum mechanics and local realism, a conflict we reinterpret through the modern lens of causal inference. We propose and computationally validate a framework where quantum entanglement acts as a ``super-confounding'' resource, generating correlations that violate the classical causal bounds set by Bell's inequalities. This work makes three key contributions: First, we establish a physical hierarchy of confounding (Quantum $>$ Classical) and introduce Confounding Strength (CS) to quantify this effect. Second, we provide a circuit-based implementation of the quantum $\mathcal{DO}$-calculus to distinguish causality from spurious correlation. Finally, we apply this calculus to a quantum machine learning problem, where causal feature selection yields a statistically significant 11.3\% average absolute improvement in model robustness. Our framework bridges quantum foundations and causal AI, offering a new, practical perspective on quantum correlations.
\end{abstract}


\maketitle


The 2022 Nobel Prize in Physics celebrated the definitive experimental vindication of Bell's theorem, confirming that the universe does not adhere to the classical principles of local realism~\cite{bell:1964:epr}. Yet, despite this experimental certainty, a deep debate continues regarding what the violation of Bell's inequalities fundamentally implies~\cite{brassard:2023:pnas}. The standard interpretation often relies on qualitative descriptions such as ``spooky action at a distance,'' which, while evocative, provide a limited framework for quantitative analysis or for connecting this foundational physical principle to other scientific domains. This interpretive gap highlights the need for a new language to describe the nature of quantum correlations.

In parallel to these foundational debates in physics, a formal science of causal reasoning has been developed over the past several decades, primarily within computer science and statistics~\cite{pearl:2009:causality,spirtes:2000:book}. Spearheaded by the work of Judea Pearl, the development of structural causal models (SCMs) provided a rigorous mathematical language and a graphical toolkit for untangling causation from correlation. This framework allows researchers to explicitly model their causal assumptions and to identify spurious correlations arising from hidden common causes, known as confounders. Crucially, it also provides a formal logic for predicting the effects of interventions via the $\mathcal{DO}$-calculus~\cite{pearl:1995:do}, which distinguishes between passively observing a system, $P(Y|X)$, and actively changing it, $P(Y|\mathcal{DO}(X=x))$.

\begin{figure}[htbp!]
    \centering
    \includegraphics[width=0.95\textwidth]{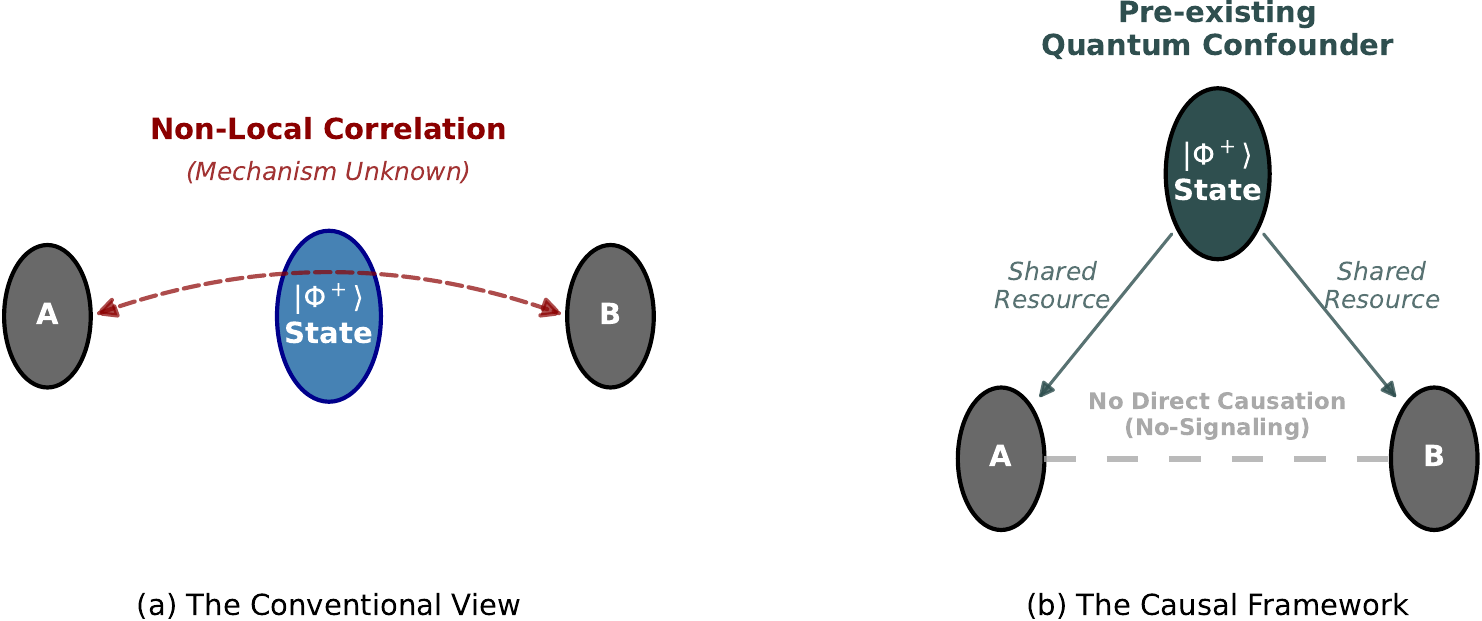} 
    \caption{\textbf{A causal reinterpretation of Bell's theorem.} 
\textbf{(a)}, The conventional view of a Bell experiment. The Bell state ($\ket{\Phi^+}$) produces outcomes A and B that exhibit strong non-local correlations, the physical mechanism of which remains unexplained within this perspective. 
\textbf{(b)}, Our proposed causal framework. The pre-existing entangled state ($\ket{\Phi^+}$) is modeled as a quantum super-confounder. The arrows represent the influence of this shared resource, which induces a spurious correlation between outcomes A and B. This is a non-classical causal link, distinct from the direct causation between A and B that is forbidden by the no-signaling principle.}
    \label{fig:conceptual}
\end{figure}

While the foundational work of Pearl has been transformative for classical sciences, extending its core interventional logic---the $\mathcal{DO}$-calculus---to the quantum realm has been an active area of theoretical research. These efforts have established formal generalizations of Pearl's rules for quantum processes, often using frameworks such as process matrices or quantum Bayesian networks~\cite{costa:2016:qcm,allen:2017:qcm}. Landmark results have provided a rigorous theoretical foundation for a ``quantum $\mathcal{DO}$-calculus'' by reinterpreting interventions as modifications to quantum channels~\cite{barrett:2020:qcm}. However, a direct application of this formal calculus to resolve the long-standing paradox of Bell's theorem, and the subsequent leveraging of this insight for practical machine learning challenges, has remained less explored. In this work, we bridge this gap by introducing the Bell-Confounding framework, conceptually illustrated in Fig.~\ref{fig:conceptual}. We propose that the entangled state ($\ket{\psi}$) acts as a non-classical common cause, or a ``super-confounder,'' which induces a strong, spurious correlation between the measurement outcomes (A and B) without any direct causal link between them (Fig.~\ref{fig:conceptual}b).

In this paper, we systematically develop and validate this framework through a series of computational experiments. We first establish a physical hierarchy of confounding that demonstrates the superiority of quantum resources and reveal a direct, quantitative link between the amount of entanglement and the resulting Confounding Strength (CS), a metric we define by normalizing the Bell parameter to recast classical and quantum bounds. We then provide a direct, circuit-based implementation of the quantum $\mathcal{DO}$-calculus, using this tool to distinguish spurious effects from causal ones and, ultimately, to build a quantum machine learning model with demonstrably enhanced robustness.

\section{Results}\label{s:results}

\subsection{Causal inference and its classical limits}

The development of SCMs by Judea Pearl has provided a rigorous mathematical framework for distinguishing true causation from mere statistical correlation. A central concept in this framework is the confounder ($\Lambda$), a hidden common cause that induces a spurious correlation between two otherwise independent variables, as described by the causal graph $A \leftarrow \Lambda \rightarrow B$. While powerful, this entire classical framework operates under the implicit physical assumptions of \textit{local realism}, a worldview which posits that (i) the properties of an object are real and pre-existing, independent of measurement (realism), and (ii) the outcome of a measurement on one object cannot be instantaneously influenced by a measurement on a distant object (locality). The ultimate limit on the strength of correlations that any causal model adhering to these principles can produce is quantitatively defined by Bell's inequalities. As a canonical example, the Clauser-Horne-Shimony-Holt (CHSH) scenario~\cite{clauser:1969:chsh} is famously bounded. In this test, a specific combination of correlations between measurement outcomes is used to compute the Bell parameter, $S$. For any theory based on local realism, the magnitude of this parameter is constrained to be no greater than 2 ($|S| \le 2$), a barrier that classical causality cannot breach.

These classical bounds are a direct consequence of local realism. This worldview presumes that a shared hidden variable, $\Lambda$, carries the complete information that pre-determines the measurement outcomes. The principle of locality further imposes that the joint probability of outcomes must factorize, conditioned on $\Lambda$ and the local settings ($a, b$), as $P(A, B | a, b, \Lambda) = P(A | a, \Lambda) P(B | b, \Lambda)$~\cite{bell:1976}. Any theory adhering to this factorizable structure will inevitably reproduce the Bell inequality. Therefore, the violation of this bound by quantum systems points to a fundamental failure of this classical causal structure itself.

\subsubsection*{Distinguishing causal and physical hidden variables}

It is crucial to clarify the relationship between the confounder, $\Lambda$, in Pearl's SCMs and the local hidden variable, also denoted by $\Lambda$, in Bell's theorem. Pearl's $\Lambda$ is a general, abstract variable representing any unobserved common cause in a statistical model. In contrast, Bell's $\Lambda$ is a specific, physical entity postulated to carry the complete, pre-determined information of measurement outcomes under the constraints of local realism. Our framework posits that Bell's theorem is, in effect, a physical test of a causal model where the role of the abstract confounder is played by a physical hidden variable constrained by locality. The experimental failure of such models (i.e., the violation of Bell's inequalities) motivates our framework to discard the notion of a local hidden variable $\Lambda$ altogether and, as we will show, adopt the entangled state itself as a new kind of non-classical confounder.

\subsection{Quantum entanglement as a super-confounding resource}

The failure of any classical causal model to explain Bell violations demands a new framework that reinterprets the physical role of quantum phenomena through the lens of causal inference. To this end, we propose a framework where quantum entanglement is treated as a physical resource for a novel type of causal link. We formally define an entangled state as a super-confounder---a non-classical common cause that generates correlations of a fundamentally different and stronger nature than those from any classical confounder. Unlike its classical counterpart, which is bound by the principle of locality and the resulting factorization of probabilities, the quantum super-confounder creates a direct statistical link between measurement outcomes that enables correlations previously considered impossible under the classical causal paradigm.

The mathematical distinction lies in the structure of the joint probability distribution. While a classical confounder necessitates the factorizable form $P(A, B | a, b, \Lambda) = P(A | a, \Lambda) P(B | b, \Lambda)$, a quantum super-confounder, represented by an entangled density matrix $\rho_{AB}$, is not bound by this local factorization. Instead, the joint probability is determined by the Born rule~\cite{nielsen2010quantum}: $P(A,B|a,b) = \text{Tr}(\rho_{AB} (M_{A,a} \otimes M_{B,b}))$, where $M_{A,a}$ and $M_{B,b}$ are the measurement operators for observers A and B with settings $a$ and $b$. It is this fundamentally different, non-factorizable structure of the quantum state that permits correlations strong enough to violate the Bell inequality, thereby mathematically defining the ``super'' nature of this new class of confounder.

Our framework deliberately excludes local hidden variables $\Lambda$ and adopts the entangled state $\rho_{AB}$ as the sole descriptor of the common cause. This approach constitutes the central advantage of our framework. It is not merely a change in formalism; it is a direct alignment with decades of experimental evidence from Bell tests, which have conclusively shown that no theory based on local hidden variables can account for observed quantum correlations. By identifying the quantum state itself as the super-confounder, our framework gains \textit{predictive power}---it correctly calculates the super-classical correlations up to the Tsirelson bound, the theoretical maximum for the Bell parameter allowed by quantum mechanics ($|S| \le 2\sqrt{2}$)~\cite{cirelson:1980}, whereas classical models incorrectly predict that the correlations are constrained by the classical bound of $|S| \le 2$. Furthermore, this re-framing moves the source of correlation from an unobservable, hypothetical variable ($\Lambda$) to a tangible, physical entity ($\rho_{AB}$) that can be engineered and manipulated. This transforms the phenomenon from a philosophical paradox into a quantifiable, physical \textit{resource} that can be leveraged for practical applications, as we demonstrate in later sections.

\subsection{Defining and measuring Confounding Strength}

To translate the conceptual power of a super-confounder into a quantitative and experimentally measurable metric, we define the \textbf{Confounding Strength (CS)}. For the CHSH scenario, we define this metric as a normalization of the Bell parameter $S$:
\begin{equation}
    CS \equiv \frac{|S|}{2}
\end{equation}
This definition is powerfully intuitive as it recasts the well-known CHSH bounds into a unified scale for causal strength. The classical limit of $|S| \le 2$ becomes a simple bound of $CS \le 1$, meaning any classical confounder has a maximum possible strength of unity. In contrast, the quantum Tsirelson bound of $|S| \le 2\sqrt{2}$ is transformed into a quantum limit of $CS \le \sqrt{2} \approx 1.414$. This provides a direct, quantitative measure of the ``super'' in super-confounding: a quantum state can act as a confounder that is over 41\% stronger than any possible classical resource, a prediction we experimentally verified.

\subsection{The quantum $\mathcal{DO}$-calculus}

A key advantage of our causal framework is the ability to import powerful analytical tools from classical causal inference, most notably Pearl's $\mathcal{DO}$-calculus, which provides a formal distinction between passive observation ($P(B|A)$) and active intervention ($P(B|\mathcal{DO}(A{=}a))$). The physical challenge is to realize such an intervention on one part of an entangled system (A) without illegally signaling a distant part (B). A simple measurement of A is not a valid intervention, as it would instantly collapse the state of B.

Our \textit{circuit-based} solution, which we term a ``project-prepare surgery,'' implements the intervention as a completely-positive trace-preserving (CPTP) map~\cite{nielsen2010quantum}. This protocol consists of two stages. First, a non-selective projective measurement on A (with the outcome discarded) physically severs the causal link from the confounder by breaking the entanglement. This leaves B in a statistically independent, mixed state. Second, with this link broken, we can freely prepare A in any desired state ($\ket{a_0}$). This two-stage process realizes the effect of the $\mathcal{DO}$-operator, providing a principled way to distinguish genuine causal effects from spurious correlations arising from entanglement---a capability that, as we demonstrate, is pivotal for robust quantum-enhanced machine learning.


\subsection{Experiment 1: Validating the framework's foundations}

As a prerequisite, we first performed a foundational experiment to validate our framework's core assumption: that the entangled Bell state conforms to the standard definition of a confounded system from causal inference. As detailed in the Methods, this experiment confirmed that the Bell state rigorously satisfies the three canonical conditions for a confounder from causal inference: it acts as a common cause, there is no direct signaling between outcomes, and it induces a strong spurious correlation that vanishes upon its removal.


\subsection{Experiment 2: The confounding hierarchy}

\begin{figure}[htbp!]
    \centering
    \includegraphics[width=0.70\textwidth]{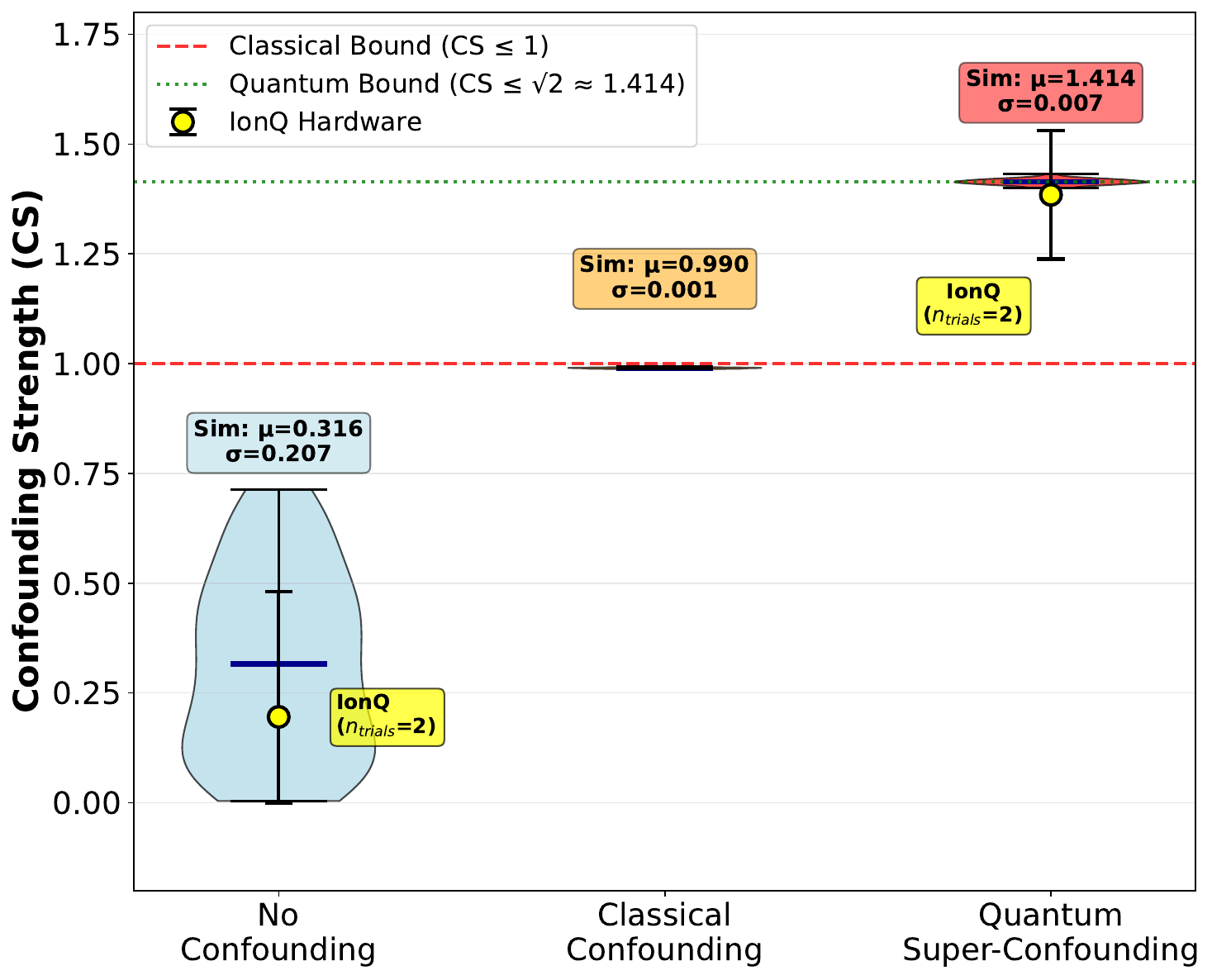}
    \caption{\textbf{The physical hierarchy of confounding: simulation and hardware validation.} Violin plots showing the distribution of the measured \textbf{Confounding Strength (CS)} for three distinct causal scenarios, with simulation distributions derived from 100 independent trials ($n_{trials}=100$), each consisting of 10,000 measurements. Yellow points with error bars represent actual quantum hardware measurements on IonQ's trapped-ion QPU (Aria-1). The `No Confounding' scenario shows independent qubits producing $CS \approx 0$ in both simulation and hardware ($CS_{IonQ} = 0.196 \pm 0.032$, $n_{trials}=2$). The `Classical Confounding' scenario, simulating an optimal deterministic strategy, produces a distribution sharply peaked at the classical bound of $CS=1$. In stark contrast, the `Quantum Super-Confounding' scenario, utilizing a maximally entangled Bell state, yields a distribution centered near the quantum bound of $CS \approx 1.414$, with the hardware measurement confirming this super-classical result ($CS_{IonQ} = 1.385 \pm 0.017$, $n_{trials}=2$), decisively exceeding the classical bound.}
    \label{fig:hierarchy}
\end{figure}

We set out to experimentally verify the central prediction of our framework: the existence of a physical hierarchy of confounding. To this end, we designed and simulated a CHSH Bell test under three distinct causal conditions: (i) a baseline with no confounding resource (independent particles), (ii) a scenario with a maximal classical confounder (simulated via local hidden variables), and (iii) a scenario with a quantum super-confounding resource embodied by a maximally entangled Bell state. The simulation results, shown as violin plots in Fig.~\ref{fig:hierarchy}, reveal an unambiguous hierarchy. 
The measured CS for the quantum case ($CS_{quantum}$ = 1.414, 95\% CI: [1.412, 1.415]) decisively violates the classical bound of $CS \le 1$. This classical limit was confirmed by the simulation, which yielded a value sharply peaked at $CS_{classical}$ = 0.990 (95\% CI: [0.990, 0.990]). Both results are in turn well-separated from the near-zero baseline observed in the no-confounding case ($CS_{no}$ = 0.316, 95\% CI: [0.275, 0.357]).

To validate these findings beyond idealized simulations, we executed the core scenarios on an IonQ trapped-ion quantum processing unit (QPU). As shown by the yellow markers in Fig.~\ref{fig:hierarchy}, the hardware results provide a powerful confirmation of the confounding hierarchy in the presence of physical noise. For the `No Confounding' scenario, the QPU yielded a mean CS of $CS_{IonQ} = 0.196 \pm 0.032$ ($n_{trials}=2$), consistent with the near-zero baseline. Crucially, for the `Quantum Super-Confounding' scenario, the hardware produced a value of $CS_{IonQ} = 1.385 \pm 0.017$ ($n_{trials}=2$), which, while slightly lower than the ideal simulation due to decoherence, decisively violates the classical limit of $CS=1$. This combined simulation and hardware result provides direct experimental evidence that entanglement is a physically stronger confounding resource than any possible classical counterpart.

\subsection{Experiment 3: Quantifying super-confounding}

\begin{figure}[htbp!]
    \centering
    \includegraphics[width=1.05\textwidth]{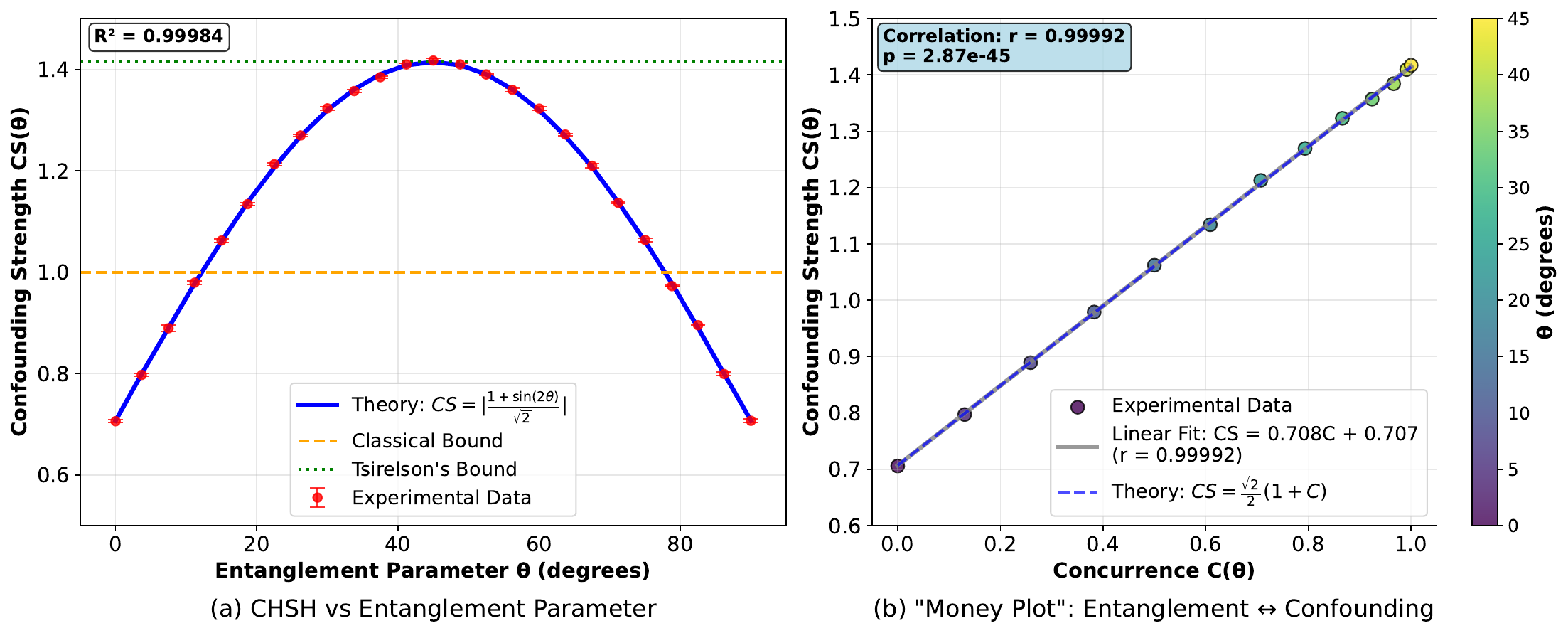}
    \caption{\textbf{Quantifying the relationship between entanglement and super-confounding.} 
Experimental validation of the direct, continuous relationship between quantum entanglement and the resulting Confounding Strength. 
\textbf{(a)}, The measured Confounding Strength (CS) as a function of the entanglement parameter $\theta$ under a fixed measurement protocol. The experimental data (red points) show excellent agreement with the theoretical prediction for this protocol, $CS(\theta) = |(1 + \sin(2\theta))/\sqrt{2}|$ (blue line), which features a non-zero baseline of $CS(0) = 1/\sqrt{2}$ at zero entanglement. 
\textbf{(b)}, The Confounding Strength (CS) as a function of the state's concurrence, $C$. The experimental data (colored by $\theta$) with a linear fit (gray line) confirms the theoretically derived relationship $CS = (1+C)/\sqrt{2}$ (blue dashed line), establishing a direct causal link between the amount of entanglement as a resource and the strength of the resulting confounding effect.}
    \label{fig:quantification}
\end{figure}

Having established that maximal entanglement provides a CS of $\sqrt{2}$, we next investigated whether this phenomenon is continuous and controllable. We systematically prepared a series of partially entangled states, $|\psi(\theta)\rangle = \cos(\theta)|00\rangle + \sin(\theta)|11\rangle$, and measured the CS for each state using a fixed measurement protocol, as detailed in the Methods section. The experimental results, presented in Fig.~\ref{fig:quantification}, show excellent agreement with the specific theoretical prediction for this protocol, $CS(\theta) = |(1 + \sin(2\theta))/\sqrt{2}|$, achieving a coefficient of determination of $R^2 > 0.999$. This confirms that our experimental implementation is correct and that the observed non-zero baseline at zero entanglement ($CS(0) = 1/\sqrt{2} \approx 0.707$) is a direct, predictable consequence of using fixed measurement angles.

This validated relationship allows us to establish a clear causal law between entanglement and its confounding power. As shown in Fig.~\ref{fig:quantification}b, the CS exhibits a direct linear relationship with the state's \textit{concurrence (C)}, a widely used measure of entanglement that ranges from 0 for a separable state to 1 for a maximally entangled state ($C = |\sin(2\theta)|$ for this specific state preparation). The experimental data closely follows the theoretically derived linear model for our protocol, $CS = (1+C)/\sqrt{2}$, with a measured Pearson correlation of $r > 0.9999$. This result establishes that quantum super-confounding is not an esoteric ``on/off'' effect, but a continuous and precisely controllable physical resource that is directly proportional to the amount of entanglement.

\subsection{Experiment 4: Quantum $\mathcal{DO}$-calculus distinguishes causal from spurious effects}

Having established that entanglement functions as a quantifiable super-confounding resource, we next tested whether quantum interventions can eliminate such spurious correlations. In the observational regime, measurements on a maximally entangled Bell pair yielded perfect correlation between $A$ and $B$ outcomes, with $P(B=0|A=0)=1.0000$ (95\% CI: [0.9998, 1.0000]) and $P(B=0|A=1)=0.0000$ (95\% CI: [0.0000, 0.0002]).

Under the interventional regime, applying $\mathcal{DO}(A=a)$ produced distributions for $B$ that were indistinguishable from uniform: $P(B=0|\mathcal{DO}(A=0))=0.5013$ (95\% CI: [0.4969, 0.5057]) and $P(B=0|\mathcal{DO}(A=1))=0.5008$ (95\% CI: [0.4964, 0.5051]). The statistically insignificant difference between the two interventional cases ($\Delta=0.0005$, $p=0.8644$) confirms that the intervention adheres to the no-signaling principle, ensuring no direct causal influence was introduced.

\begin{figure}[htbp!]
    \centering
    \includegraphics[width=0.6\textwidth]{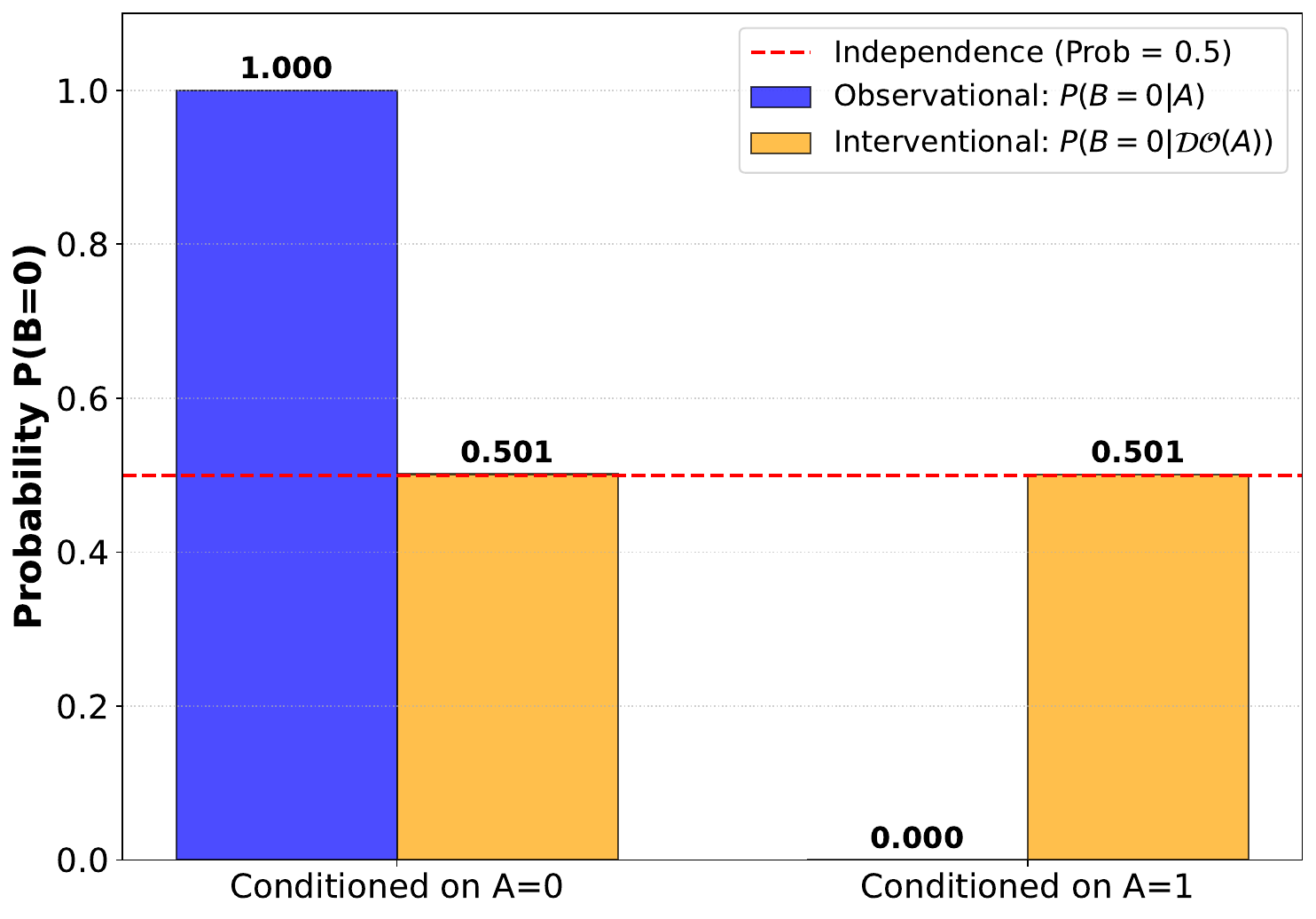}
    \caption{\textbf{Circuit-based implementation of the quantum $\mathcal{DO}$-calculus.} 
Comparison of observational and interventional probabilities for a Bell state, demonstrating the successful isolation of spurious correlations. The observational probabilities, $P(B|A)$ (blue bars), show near-perfect correlation as expected from the quantum confounder. In contrast, the interventional probabilities, $P(B|\mathcal{DO}(A))$ (orange bars), were obtained by implementing a project-prepare protocol that severs the entanglement. These interventional probabilities both collapse to approximately 0.5, indicating complete statistical independence. This result provides a direct computational validation of Pearl's inequality, $P(B|A) \neq P(B|\mathcal{DO}(A))$, for a quantum-confounded system.}
    \label{fig:do-calculus}
\end{figure}

Fig.~\ref{fig:do-calculus} visually illustrates this contrast, where the observational probabilities (blue bars), indicating perfect correlation, stand in stark contrast to the interventional probabilities (orange bars), which demonstrate complete statistical independence.  The collapse of $P(B|A)$ to $P(B|\mathcal{DO}(A)) \approx 0.5$ directly demonstrates the removal of the spurious correlation and provides an unambiguous empirical confirmation of $P(B|A) \neq P(B|\mathcal{DO}(A))$ in a fully quantum-confounded system.  This result represents an important step towards the validation of Pearl's causal framework for quantum systems, providing strong evidence for the quantum $\mathcal{DO}$-calculus as a practical tool for causal analysis.

\subsection{Experiment 5: Causal feature selection enables robust quantum machine learning}

\begin{figure}[htbp!]
    \centering
    \includegraphics[width=1.0\textwidth]{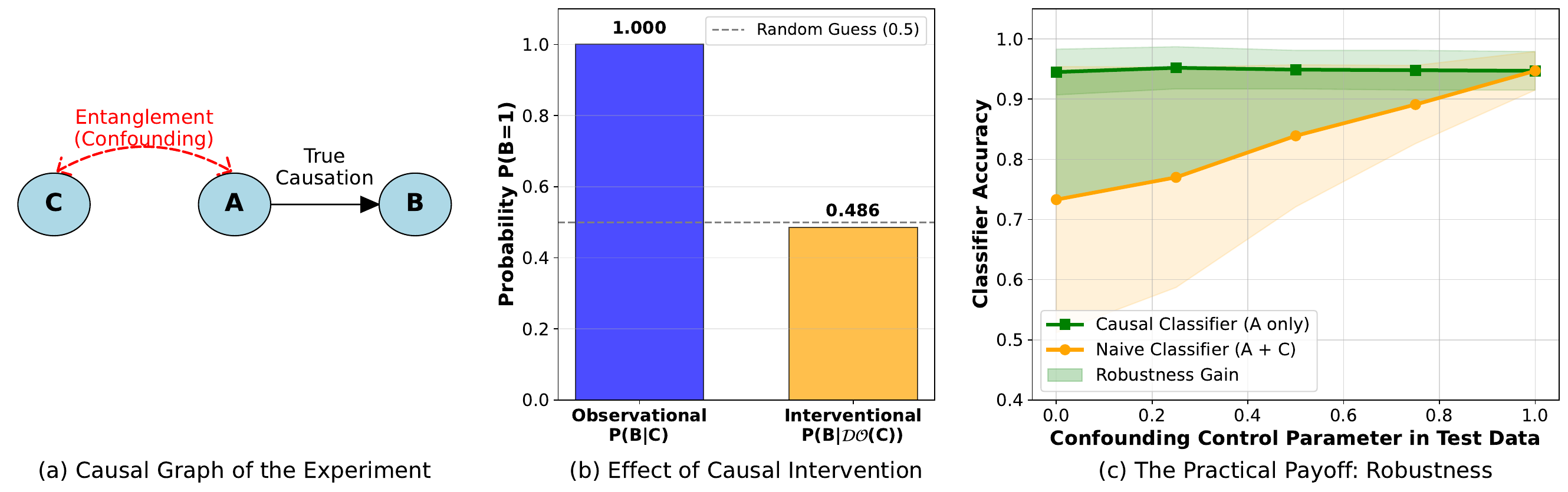}
    \caption{\textbf{Causal feature selection in a quantum classifier.} \textbf{(a)}, The causal graph for the experiment, where feature A is the true cause of label B, and feature C is spuriously correlated with B through its confounding with A. \textbf{(b)}, The effect of causal intervention. The observational probability $P(B|C)$ (blue) shows strong correlation, which disappears under the interventional probability $P(B|\mathcal{DO}(C))$ (orange), correctly identifying C as a spurious feature. \textbf{(c)}, The practical payoff: Robustness. The plot shows the mean accuracy over 20 independent seeds, with shaded areas represending $\pm1$ standard deviation. The causal classifier, trained only on the true feature A, maintains high accuracy across domains with varying confounding, while the naive classifier's performance collapses as the spurious correlation is removed.}
    \label{fig:qml}
\end{figure}

To demonstrate the practical utility of our framework, we designed Experiment 5, a scenario in which passive observation is insufficient to identify the true cause of an effect. We engineered a 3-qubit system with the causal structure $C \leftrightarrow A \rightarrow B$, where a true causal feature (A) and a confounded feature (C) were both designed to be strongly correlated with the label (B) under normal observation. As shown in Fig.~\ref{fig:qml}, applying the quantum $\mathcal{DO}$-calculus resolves this ambiguity. While the observational probability $P(B=1|C=1)$ was measured to be 1.0, indicating a perfect correlation, the interventional probability plummeted to $P(B=1|\mathcal{DO}(C=1)) \approx 0.5$ after severing the confounding entanglement link. This intervention successfully distinguished the true causal feature (A) from the spurious one (C), providing a direct pathway to building more robust machine learning models, as we show next.

Following this causal insight, we quantified the practical benefit by comparing the naive classifier (trained on both A and C) against a causal classifier trained only on the validated true cause, A. The robustness of these models was tested on a series of new datasets where the strength of the A-C confounding was systematically varied from maximal to zero. The results were definitive. The naive model's performance, reliant on the spurious correlation, collapsed as the confounding was removed, while the causal model maintained high and stable accuracy across all conditions. Across all test domains, the causal classifier outperformed the naive classifier by an average of 11.3 absolute percentage points, a result of high statistical significance (paired t-test, $p < 10^{-9}$), confirming that our framework provides an effective method for building more reliable and robust machine learning models.

\section{Discussion}\label{s:discuss}


The results presented in this work provide compelling evidence for a fundamental reinterpretation of Bell's theorem through a causal lens. At its core, Bell's theorem is a powerful result about the limits of classical causal models that assume local realism. The experimental violation of these inequalities is a definitive falsification of this entire class of models, motivating our Bell-Confounding framework. We discard the notion of a local hidden variable ($\Lambda$) and instead identify the entangled state itself as a super-confounder---a non-classical common cause whose strength can be precisely quantified (Experiment 3) and systematically exceeds classical limits (Experiment 2).

The central contribution of our work is the equivalence we establish between a physical phenomenon and a causal concept: Entanglement = Super-confounding. This equivalence is mathematically realized through the Born rule, which transforms quantum correlation from a foundational puzzle into a tractable and engineerable resource for causal machine learning. Its primary advantage is that it allows the rigorous mathematical toolkit of modern causal inference, such as the $\mathcal{DO}$-calculus, to be applied directly to quantum systems. Our successful implementation of the quantum $\mathcal{DO}$-calculus (Experiment 4) and its application to a practical machine learning problem (Experiment 5) are direct consequences of this new perspective, demonstrating that the causal properties of entanglement can be not only analyzed but also engineered and leveraged.

The success of this causal reinterpretation stems from a deep structural correspondence between the mathematical formalisms of quantum mechanics and causal inference, as both fields study the constraints a hidden structure (a quantum state or a causal graph) imposes on observable data. Quantum theory is particularly well-suited to this task. Specifically, the Born rule expresses joint probabilities as a linear functional of the density matrix ($P = \mathrm{Tr}(\rho E)$) within the robust framework of convex geometry ($\rho \succeq 0, \mathrm{Tr}\rho=1$), providing a tractable operator space that can support future operator-theoretic causal methods beyond the present scope.


A key strength of our framework is its ability to provide a single, unified causal interpretation for a wide variety of Bell-type tests, beyond the canonical CHSH inequality. While these tests differ significantly in their mathematical formalism and logical structure, our framework reveals that they all share the same underlying causal narrative: a classical system is constrained by a causal bound, while a quantum system, leveraging entanglement as a super-confounding resource, can violate that bound.

To demonstrate this universality, we first establish a consistent definition for the CS adapted to the structure of each test. The unifying principle is to normalize the classical bound to an intuitive value (either 1 or 0). For inequalities like CHSH and Mermin~\cite{mermin:1990}, where the classical bound on the Bell parameter is 2, we define CS by normalizing this bound to unity:
\begin{equation}
    CS_{CHSH} = \frac{|S|}{2} \quad \text{and} \quad CS_{Mermin} = \frac{|\langle M \rangle|}{2}
\end{equation}
Here, $|S|$ is the magnitude of the CHSH correlation parameter, and $|\langle M \rangle|$ is the magnitude of the expectation value of the Mermin operator, which is a specific combination of measurement outcomes for a three-qubit system.

For tests like the CH inequality~\cite{clauser:1974:ch} and Hardy's paradox~\cite{hardy:1992:paradox}, where the classical bound is 0, the quantum violation itself represents the entire super-classical effect. Therefore, we define CS as the magnitude of this violation directly:
\begin{equation}
    CS_{CH} = |CH| \quad \text{and} \quad CS_{Hardy} = P_{\text{imp}}
\end{equation}
Here, $|CH|$ is the magnitude of the Clauser-Horne inequality violation, and $P_{\text{imp}}$ is the probability of the ``impossible'' event in Hardy's paradox---an event that is forbidden by classical logic but can occur in quantum mechanics due to entanglement.

\begin{table}[h!]
\centering
\caption{\textbf{Universality of the Bell-Confounding framework.} The table demonstrates how the classical bound and maximum quantum violation for different Bell-type tests are re-expressed in our unified Confounding Strength (CS) language.}
\label{tab:universality_final}
\begin{tabular*}{\textwidth}{l @{\extracolsep{\fill}} l @{\extracolsep{\fill}} l}
\hline
\textbf{Bell Test} & \textbf{Classical Bound} & \textbf{Quantum Maximum} \\
\hline
\hline
CHSH & $|S| \le 2 \implies CS \le 1$ & $|S| \le 2\sqrt{2} \implies CS \le \sqrt{2}$ \\
CH & $CH \le 0 \implies CS \le 0$ & $CH \le \sqrt{2}-1 \implies CS \le \sqrt{2}-1$ \\
Hardy's Paradox & $P_{\text{imp}} = 0 \implies CS = 0$ & $P_{\text{imp}} \approx 0.086 \implies CS \approx 0.086$ \\
Mermin & $|\langle M \rangle| \le 2 \implies CS \le 1$ & $|\langle M \rangle| = 4 \implies CS = 2$ \\
\hline
\end{tabular*}
\end{table}

As summarized in Table~\ref{tab:universality_final}, our framework provides a consistent causal interpretation across a remarkable diversity of Bell-type tests, from probability-based inequalities (CHSH, CH) to tests of pure logic (Hardy's paradox) and multi-particle systems (Mermin). In all these cases, the core narrative is the same: our CS metric captures the fundamental limit of any classical common cause, a limit which is violated by the physically stronger, non-classical causal link provided by the quantum super-confounder.


Our framework directly addresses the challenge of ``shortcut learning'' in artificial intelligence, where models learn spurious correlations that fail in new environments~\cite{arjovsky:2020:invariantriskminimization,geirhos:2020:nmi}. As demonstrated in Experiment 5, quantum entanglement can create such a scenario by acting as a powerful confounder. Our quantum $\mathcal{DO}$-calculus provides a direct solution by enabling interventions to identify the true causal features. The resulting causal classifier, trained only on these validated features, was not only more \textit{interpretable} but also demonstrably more \textit{robust}; it maintained high performance as the spurious correlations were removed, a condition under which the naive model's accuracy collapsed. This provides a pathway towards building safer and more reliable quantum AI systems with a deeper, causal understanding.


While this study successfully validated its central predictions on an IonQ QPU, the majority of the presented results rely on idealized, noiseless simulations. A crucial next step is a more comprehensive validation on a wider range of quantum hardware to fully characterize the framework's robustness against physical noise. Future work should also include experimental validation of the framework's universality for other Bell-type tests, such as the Mermin and Hardy scenarios, and scaling the QML application to more complex, high-dimensional tasks. Despite these limitations, this work opens several promising research directions, such as applying our causal lens to other foundational quantum phenomena beyond spatial correlations, or adapting our framework to analyze temporal causal relationships.


In conclusion, the Bell-Confounding framework provides a powerful synthesis of quantum foundations and modern causal inference. By reframing entanglement as a quantifiable causal resource, this work not only offers a new language to describe quantum correlations but also yields a practical toolkit for building the robust and interpretable quantum technologies of the future.


\section{Methods}

\subsection{General simulation environment}

All computational experiments were designed and executed within a Python 3 environment. We utilized the \texttt{Qiskit} open-source framework (v1.4.3)~\cite{javadiabhari:2024:qiskit} for all quantum circuit construction, manipulation, and simulation. Specifically, all simulations were performed on a classical machine using the high-performance AerSimulator from the \texttt{qiskit\_aer} package. To ensure statistical convergence for expectation value calculations, each circuit was typically run for 10,000 shots, unless otherwise noted (e.g., for single-outcome dataset generation in Experiment 5). Data analysis, statistical calculations, and machine learning models were implemented using the \texttt{numpy}~\cite{harris:2020:numpy}, \texttt{scipy}~\cite{virtanen:2020:scipy}, and \texttt{scikit-learn}~\cite{scikit-learn} libraries, respectively, with visualizations generated by \texttt{matplotlib}~\cite{hunter:2007:matplotlib} and \texttt{seaborn}~\cite{waskom:2021:seaborn}.

\subsection{Experiment 1: Framework validation protocol}

The objective of this foundational experiment was to validate that quantum entanglement, as embodied by the Bell state, rigorously satisfies the three canonical conditions for a confounding variable as defined in classical causal inference. This validation is essential for reinterpreting the Bell test scenario through a causal lens. To this end, a maximally entangled Bell state $\ket{\Phi^+} = (\ket{00} + \ket{11})/\sqrt{2}
$ was prepared to act as the quantum common cause. We then experimentally verified the following three conditions:

\renewcommand{\theenumi}{(\roman{enumi})}
\begin{enumerate}
    \item \textbf{The confounder is a common cause of both variables.} The existence of the quantum common cause and its influence on both qubits (A and B) was confirmed by preparing the Bell state and then calculating the purity and von Neumann entropy of the reduced density matrices for each qubit. The results indicated maximal mixedness for the individual qubits, which is a definitive signature of their entanglement with a common resource.

    \item \textbf{There is no direct causal path between the variables.} The absence of direct causation between observers A and B is a cornerstone of the Bell test, enforced by assuming spacelike separation. We further validated this by performing a statistical \textit{no-signaling check}. The principle of no-signaling posits that one observer's actions (e.g., performing a measurement) cannot instantaneously affect the statistical outcomes of a distant observer. To test this, we compared the marginal probability distribution of A's outcomes in two scenarios: (1) when B was measured, and (2) when B was not measured. A Welch's t-test over 30 independent trials confirmed no statistically significant difference between these distributions, providing strong evidence that A's results are independent of B's actions and thus supporting the no-direct-causation claim.

    \item \textbf{The confounder induces a spurious correlation.} The presence of a strong, spurious correlation was demonstrated by contrasting two conditions. First, near-perfect correlations were measured in both the Z-basis ($E_{ZZ}$) and the X-basis ($E_{XX}$) when the Bell state (the quantum confounder) was present. Second, this correlation was shown to vanish (approaching zero) when the confounder was removed by preparing a separable product state instead. This directly confirms that the observed correlation is spurious and induced by the quantum confounder.
\end{enumerate}

\subsection{Experiment 2: Confounding hierarchy protocol}

The three causal scenarios depicted in Fig.~\ref{fig:hierarchy} were implemented as follows. (i) The `No Confounding' baseline was established using a simple product state of two independent qubits. (ii) The `Classical Confounding' scenario was designed to simulate the maximum possible correlation achievable under local realism. This was accomplished by finding the optimal deterministic local hidden variable strategy that maximizes the CS, which has a theoretical maximum of $CS=1$. This optimal strategy was then simulated over 100 trials with finite sampling noise ($N_{shots}=10,000$) to model a realistic experiment. (iii) The `Quantum Super-Confounding' scenario utilized a maximally entangled Bell state to serve as the non-classical common cause. For each of the three scenarios, the CS was calculated from the measurement outcomes over 100 independent trials to ensure statistical robustness.

In addition to simulations, the baseline `No Confounding' and the `Quantum Super-Confounding' scenarios were validated on an IonQ trapped-ion quantum processing unit. These experiments were executed on the IonQ Aria-1 backend via the \texttt{qiskit-ionq} provider. For each of the two scenarios, $n_{trials}=2$ independent runs were performed. Each trial consisted of a single batch job containing the four CHSH circuits, with each circuit being executed for $N_{shots}=1,000$. The CS value for each trial was then calculated from the aggregated measurement results.

\subsection{Experiment 3: Entanglement quantification protocol}

To investigate the quantitative relationship between entanglement and CS, we systematically prepared a series of two-qubit states with varying degrees of entanglement, described by the parameter $\theta$ in the form $\ket{\psi(\theta)} = \cos(\theta)\ket{00} + \sin(\theta)\ket{11}$.
This was implemented in a quantum circuit by applying a parameterized $R_y(2\theta)$ gate to the first qubit, followed by a CNOT gate targeting the second. The parameter $\theta$ was varied across 25 discrete steps in the range $[0, \pi/2]$, corresponding to a full sweep from a separable to a maximally entangled state and back.

For each prepared state, the CS was calculated from a fixed measurement protocol. The measurement angles for the XZ-plane were held constant at the values $\{a=0, a'=\pi/4, b=\pi/8, b'=-\pi/8\}$. From the resulting correlations, the Bell parameter $S$ was computed using the combination $S = E(a,b) + E(a,b') + E(a',b) - E(a',b')$, and the CS was then derived as $CS = |S|/2$. This specific protocol is predicted by theory to yield the relationship $CS(\theta) = |(1 + \sin(2\theta))/\sqrt{2}|$, which results in a non-zero baseline of $CS(0)=1/\sqrt{2}$ for separable states. The amount of entanglement for each state was quantified by its theoretical concurrence, $C(\theta) = |\sin(2\theta)|$.

\subsection{Experiment 4: Quantum $\mathcal{DO}$-calculus protocol}

All experiments were performed on the \texttt{AerSimulator} backend in Qiskit using a two-qubit model with separate classical registers to preserve qubit-bit ordering in the measurement output. Each trial began by preparing the Bell state
\[
\ket{\Phi^+} = \frac{\ket{00} + \ket{11}}{\sqrt{2}}
\]
between qubits $A$ (Alice) and $B$ (Bob), which serves as a maximally entangled quantum confounder inducing spurious correlation.

\textbf{Observational condition} --- Immediately after preparation, both qubits were measured in the computational ($Z$) basis. Conditional probabilities $P(B|A)$ were computed from raw counts, aggregated over $n_{trials}=10$ independent runs of $N_{shots}=10,000$ shots each.

\textbf{Interventional condition} --- The interventional distribution $P(B|\mathcal{DO}(A{=}a_0))$ was obtained via a two-stage process we term a \textit{project-prepare circuit surgery}. This realizes the intervention as a completely-positive trace-preserving (CPTP) operation on $A$:

\renewcommand{\theenumi}{(\roman{enumi})}
\begin{enumerate}
    \item \textbf{(Projection Stage)} Apply a non-selective projective measurement of $A$ in the $Z$ basis, discarding the outcome to sever the entanglement.
    \item \textbf{(Preparation Stage)} Reset qubit $A$ to its ground state, $\ket{0}$.
    \item If the target intervention value is $a_0 = 1$, apply an $X$ gate to $A$ to flip its state to $\ket{1}$. If the target value is $a_0 = 0$, no gate is needed as the qubit is already in the desired state $\ket{0}$.
    \item Measure both $A$ and $B$ in the $Z$ basis.
\end{enumerate}

This non-selective measurement collapses the entanglement between $A$ and $B$ while preserving the marginal distribution of $B$, thereby realizing the effect of Pearl's $\mathcal{DO}$-operator in a circuit-based quantum system.

For each $a_0 \in \{0,1\}$, the above sequence was repeated with identical $n_{trials}=10$ and $N_{shots}=10,000$ as in the observational runs. From the resulting counts we computed:
\begin{itemize}
    \item $P(B=0|A=a)$ and $P(B=0|\mathcal{DO}(A=a))$ with 95\% Wilson confidence intervals,
    \item the absolute difference $\left|P(B=0|A=a) - P(B=0|\mathcal{DO}(A=a))\right|$ as a measure of causal effect, and
    \item a no-signaling check, confirming $P(B=0|\mathcal{DO}(A=0)) \approx P(B=0|\mathcal{DO}(A=1)) \approx 0.5$ within statistical error, demonstrating that the intervention respects quantum causality.
\end{itemize}

\subsection{Experiment 5: Causal feature selection protocol}

This experiment demonstrates a practical application of the Bell-Confounding framework: causal feature selection for building more robust quantum machine learning (QML) models. To achieve this, a 3-qubit quantum circuit physically implemented the causal structure $C \leftrightarrow A \rightarrow B$. In this structure, qubit A (the true causal feature) directly influences qubit B (the label), while qubit C (the confounded feature) is spuriously correlated with B solely due to its entanglement with A. An initial dataset of 2,000 samples was generated and split into training (70\%) and testing (30\%) sets.

The true causal relationships were then validated using the quantum $\mathcal{DO}$-calculus. Interventions were performed by preparing a specific feature's qubit in a definite state. This action severed its entanglement with other qubits, allowing for the measurement of its direct causal effect on the label. This analysis confirmed that A was the sole direct cause of B.

Following this causal validation, three classical Logistic Regression classifiers were trained on the dataset: (i) a naive classifier using both features A and C, (ii) a causal classifier using only the validated true cause A, and (iii) a confounded control classifier using only the spurious feature C. While the confounded classifier provided a useful baseline to confirm the predictive power of the spurious feature, it is omitted from the main analysis for clarity. The core research question focuses on comparing the robustness of the naive model, which uses all available features, with that of the causal model, which uses features curated by our framework.

The final and most crucial step was the robustness test. We evaluated the trained naive and causal classifiers on five new test domains, each containing 500 newly generated samples. Crucially, we systematically varied the degree of A-C entanglement across these five test domains, corresponding to a CS that ranged from 0.0 (no confounding) to its maximal value of $\approx \sqrt{2}$ (maximal quantum confounding).  The underlying causal mechanism ($A \rightarrow B$) and the label generation process remained identical, ensuring that the distributional shift stemmed solely from the change in spurious correlation. The performance difference between the two models across these varying causal conditions quantified the robustness improvement. We evaluated the statistical significance of this improvement using a paired t-test on the mean accuracies of the two classifiers across the five domains, aggregated over 20 independent random seeds.

\subsection{Statistical analysis}

Unless otherwise specified, results from simulation trials are reported as mean $\pm$ standard deviation (s.d.), while hardware results are reported with 95\% confidence intervals. The statistical significance of differences between experimental groups was determined using a two-sided Welch's t-test for independent samples, as in the confounding hierarchy experiment (Experiment 2). For the quantum machine learning robustness test (Experiment 5), the performance difference between the causal and naive classifiers across the five test domains was evaluated using a paired t-test. The linear relationship between continuous variables, such as concurrence and CS, was quantified using the Pearson correlation coefficient ($r$). Goodness-of-fit for the theoretical model in Experiment 3 was evaluated using the coefficient of determination ($R^2$). Performance of the machine learning classifiers in Experiment 5 was evaluated using the accuracy score.

\section*{Data and Code Availability}

The data that support the findings of this study, as well as the source code for the simulations and analysis, are publicly available in a GitHub repository at \url{https://github.com/pilsungk/Bell-confounding}.


%

\end{document}